\documentclass[conference]{IEEEtran}
\IEEEoverridecommandlockouts
\usepackage{cite}
\usepackage{amsmath,amssymb,amsfonts}
\usepackage{algorithmic}
\usepackage{graphicx}
\usepackage{textcomp}
\usepackage{xcolor}
\usepackage{multirow}

\def\BibTeX{{\rm B\kern-.05em{\sc i\kern-.025em b}\kern-.08em
    T\kern-.1667em\lower.7ex\hbox{E}\kern-.125emX}}
\begin{document}

\title{Decoding Event-related Potential from Ear-EEG Signals based on Ensemble Convolutional Neural Networks
in Ambulatory Environment
\thanks{This work was partly supported by Institute for Information \& Communications Technology Planning \& Evaluation (IITP) grant funded by the Korea government (MSIT) (No. 2017-0-00451, Development of BCI based Brain and Cognitive Computing Technology for Recognizing User’s Intentions using Deep Learning), (No. 2015-0-00185, Development of Intelligent Pattern Recognition Softwares for Ambulatory Brain Computer Interface), and (No. 2019-0-00079, Artificial Intelligence Graduate School Program (Korea University)). }
}

\author{\IEEEauthorblockN{Young-Eun Lee}
\IEEEauthorblockA{\textit{Dept. Brain and Cognitive Engineering} \\
\textit{Korea University}\\
Seoul, Republic of Korea \\
ye\_lee@korea.ac.kr}
\and
\IEEEauthorblockN{Seong-Whan Lee }
\IEEEauthorblockA{\textit{Dept. Artificial Intelligence} \\
\textit{Korea University}\\
Seoul, Republic of Korea \\
sw.lee@korea.ac.kr}
}

\maketitle

\begin{abstract}
Recently, practical brain-computer interface is actively carried out, especially, in an ambulatory environment. However, the electroencephalography (EEG) signals are distorted by movement artifacts and electromyography signals when users are moving, which make hard to recognize human intention. In addition, as hardware issues are also challenging, ear-EEG has been developed for practical brain-computer interface and has been widely used. In this paper, we proposed ensemble-based convolutional neural networks in ambulatory environment and analyzed the visual event-related potential responses in scalp- and ear-EEG in terms of statistical analysis and brain-computer interface performance. The brain-computer interface performance deteriorated as 3--14\% when walking fast at 1.6 m/s. The proposed methods showed 0.728 in average of the area under the curve. The proposed method shows robust to the ambulatory environment and imbalanced data as well.

\end{abstract}

\begin{IEEEkeywords}
brain-computer interface, ambulatory environment, ear-EEG, event-related potential, ensemble CNN
\end{IEEEkeywords}

\section{Introduction}
Brain-computer interfaces (BCIs) in ambulatory environment are one of the most important consideration in real life. Although many researchers have studied BCI to recognize human cognitive state or intention based on brain signals such as electroencephalography (EEG), there are limitations in processing brain signals in ambulatory environments \cite{luu2017real,luu2016unscented}. 
Artifacts in the ambulatory environment distort the brain signals, lowering the accuracy and signal-to-noise ratio (SNR) of critical components, including human intention  \cite{castermans2011optimizing,kwak2015lower,lee2020real}.
Therefore, they have attempted to decode human intention in the ambulatory environment using movement artifact removal methods \cite{gramann2010visual, bulea2015prefrontal, nordin2018dual,lee2020real} and deep neural networks \cite{kwak2017convolutional,nordin2018dual,lee2017network,kwon2019subject,lee2020decoding}.

Development of EEG measuring devices that can be used in real life to measure EEG signals have been actively conducted \cite{bleichner2017concealed,kwak2019error}.
Among them, ear-EEG has recently been extensively investigated by many researchers to increase user convenience, and has been verified by analyzing signal quality and implementing various BCI paradigms \cite{bleichner2017concealed}.
In addition, conventional scalp-EEG annoys users due to the high cost and difficulty of setting, such as using conductive gel for hair that needs to be washed after measurement and wearing a prominent cap \cite{kwak2019error}.
In order to reduce the annoyance, simple and convenient devices were designed, such as Emotive EPOC and ear-EEG. 
Kidmose et al. \cite{kidmose2013study} analyzed the scalp and ear-EEG signals using steady-state and transient event-related potential (ERP) paradigms.
In addition, Debener et al. \cite{debener2015unobtrusive,bleichner2017concealed} designed cEEGrid to placed electrodes around the ear, which can preserve the ERP components and have a similar performance to scalp-EEG. 
On the other hand, due to the limitation of lower performance than conventional scalp-EEG, several studies have attempted to increase the performance of ear-EEG for visual or auditory responses \cite{kwak2019error}.

The BCI paradigms was developed primarily with motor imagery \cite{suk2011subject,schirrmeister2017deep,kim2014decoding,lee2015subject}, ERP \cite{yeom2014efficient,chen2016high,won2017motion,lee2018high,Lee2018mental}, and steady-state visual evoked potential (SSVEP) \cite{kwak2015lower,kwak2017convolutional,Lee2018mental,muller2006steady,won2015effect}. 
ERP and SSVEP are widely used in recognizing human intentions because they have a relatively large pattern of EEG signals as a visual response and show reliable performance in terms of accuracy and response time with a small number of EEG channels compared to other BCI paradigms \cite{lee2018high,kwak2015lower}.
ERP is a time-locked brain response to certain types of stimuli (i.e. visual, auditory, haptic, etc.), with a particularly strong positive peak response, called P300, 300 msec after the appearance of a stimulus.
In particular, it is mainly used because it rarely causes eye fatigue, low BCI illiteracy, and relatively high accuracy compared to other stimuli-based paradigms \cite{muller2006steady}. 
ERP evaluation proceeds with area under the curve (AUC) rather than accuracy because the ratio of each class is different, and SNR for signal quality evaluation proceeds.

\begin{figure*}[ht]
\centering
    \includegraphics{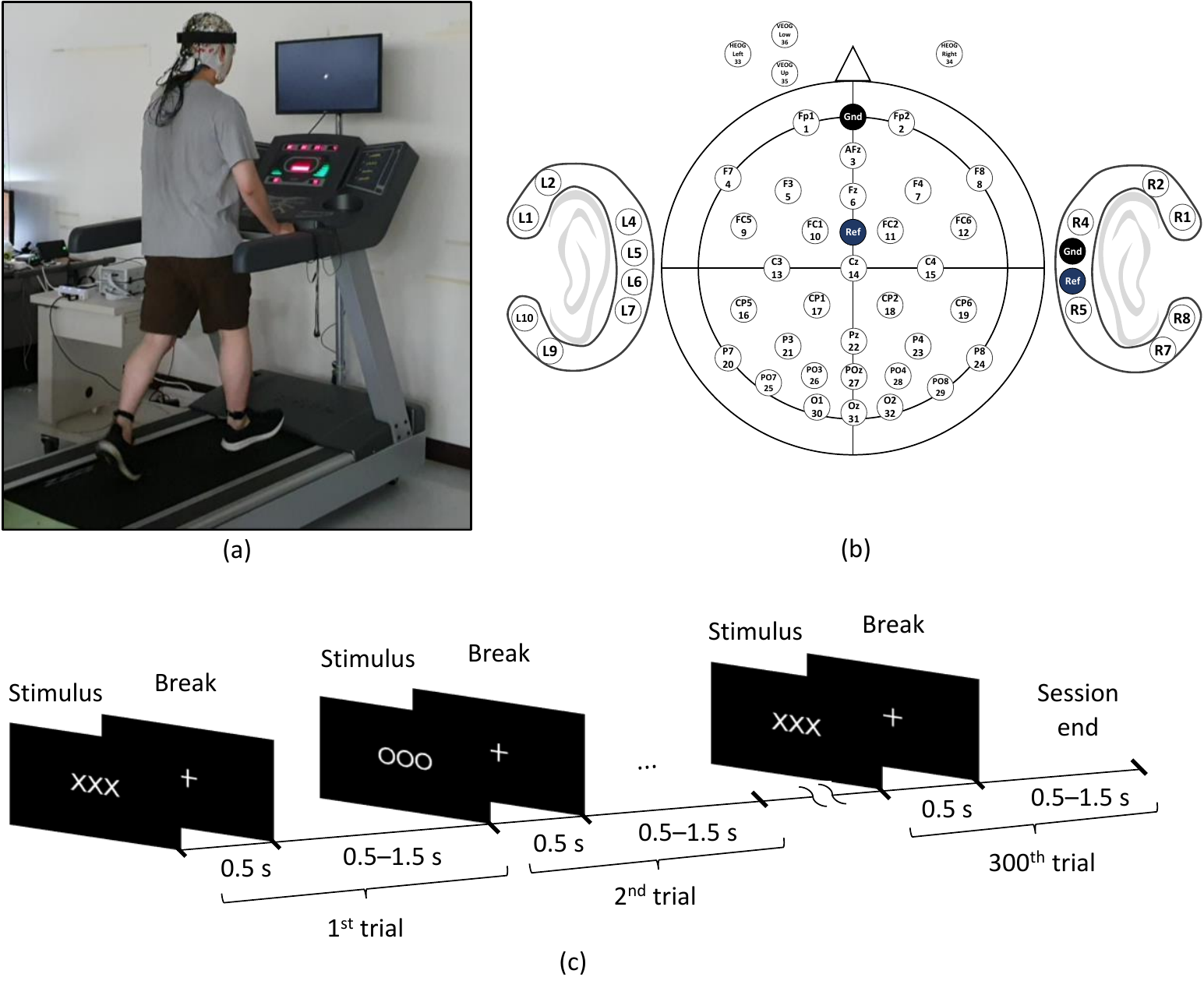}
    \caption{Design of the experiment. (a) Experimental setup and recorders. Subjects walked on a treadmill and watched a display presenting stimuli of BCI paradigms. Signals from scalp-EEG, ear-EEG around both ears, EOGs placed above and below the left eye to measure the vertical EOGs (VEOGs) and on left and right temples to measure the horizontal EOGs (HEOGs), and IMU sensors comprising nine channels, including three-axis accelerometers, three-axis gyroscopes, and three-axis magnetometers, placed on the forehead, and left and right ankles were recorded during the experiments. (c) ERP paradigm consisting of 300 trials in a session, which presents target `OOO' or non-target `XXX' for 0.5\,s, and takes a random rest for 0.5--1.5\,s. Target and non-target appear in random order, with the target ratio of 0.2. }
    \label{fig1}
\end{figure*}

There have been several attempts to use machine learning methods to recognize human intentions from visual responses.
Castermans et al. \cite{castermans2011optimizing} classified ERP intentions in the ambulatory environment up to 1.25 m/s using linear discriminant analysis classifier. Moreover, deep neural networks consisting of convolutional neural networks (CNNs) has been proposed to detect the ERP responses  \cite{lawhern2018eegnet}.
However, ERP responses typically consist of different number of trials, with different ratios of target and non-target stimuli. Thus, training with a normal deep neural network results in lower AUC by increasing the number of predictions towards non-targets.

In this paper, we decoded the ERP responses from cap-EEG and ear-EEG in the ambulatory environment. For practical BCIs, simple hardware and high accurate classifier of human intention is necessary. To solve the imbalanced problem of target and non-target ratio in ERP paradigm, we used ensemble architecture by combining results from the separated data of non-target\cite{tao2019self,yang2019hybrid,wang2020deep}. Therefore, we investigated ensemble-based comvolutional neural networks to increase human intention recognition and used ear-EEG for practical BCI in the real-world.

\begin{figure*}[ht]
    \centering
    \includegraphics{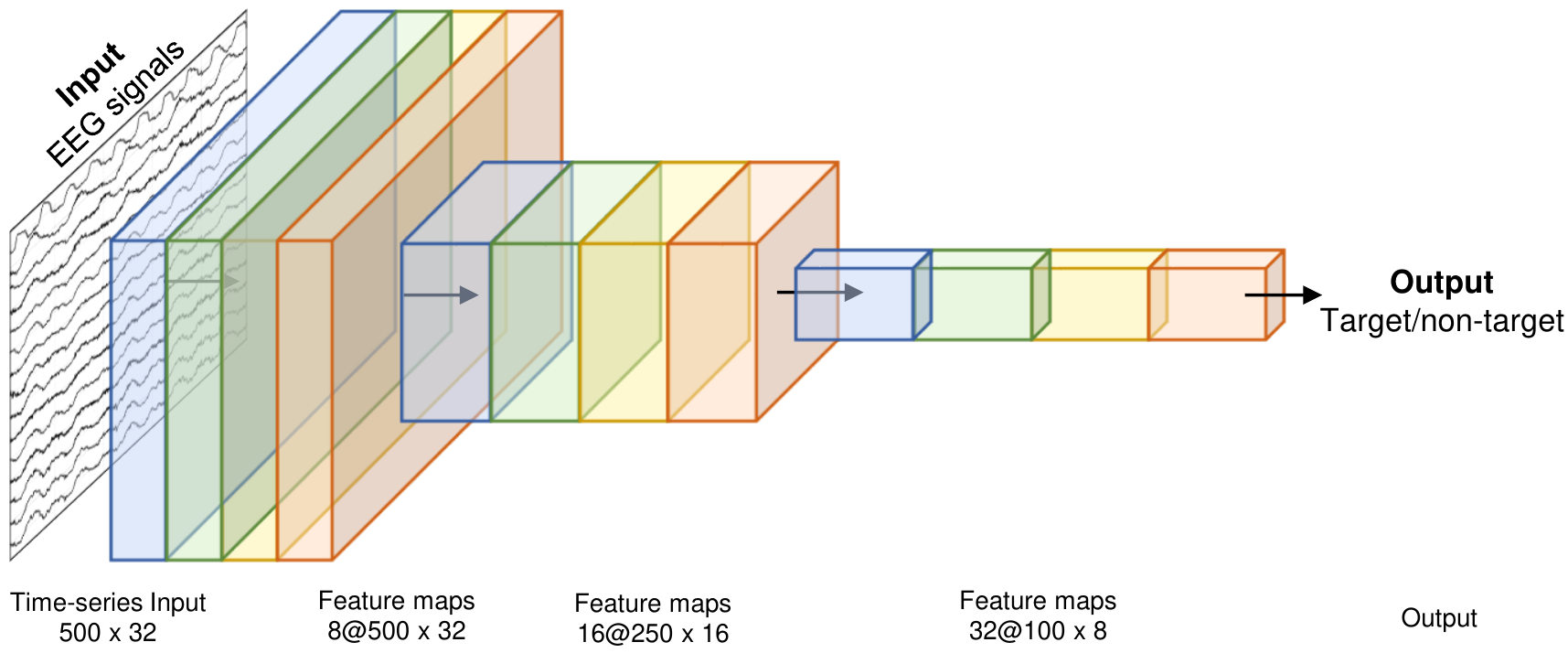}
    \caption{Proposed method architecture. The EEG signals are the input of the architecture, and the output is the classes, target or non-target.
    Each colored block indicates each ensemble network module.}
    \label{fig2}
\end{figure*}

\section{Materials and Methods}

\subsection{Participants}
We included fifteen healthy participants (2 females, age 24.8 $\pm$ 3.4 years) with normal or corrected-to-normal vision and no difficulties to walk at Korea University in Seoul, Korea. None of the participants had a history of neurological, psychiatric, or any other pertinent disease that otherwise might have affected the experimental results. This study was reviewed and approved by the Korea University Institutional Review Board (KUIRB-2019-0194-01).

\subsection{Experimental Paradigm}
The subjects were stood on the treadmill at 80 (±5) cm in front of a 60 Hz LCD monitor (Samsung, SyncMaster 2494HM, refresh rate: 60 Hz; resolution: 1920 $\times$ 1080) and walked at two different speeds (0.8 and 1.6 m/s). We experimented with target (‘OOO’) and non-target (‘XXX’) stimuli ERP paradigm with the target ratio of 0.2. The number of total trials was 300, including 60 target trials. The stimuli presented for 0.5 s and the rest time was randomly 0.5--1.5 s. The visual stimuli were generated using the Psychophysics Toolbox in Matlab. We followed the ERP paradigm of a previous study \cite{lee2019eeg}.

\subsection{Data Acquisition and Preprocessing}
Fig. \ref{fig1} shows the experimental setup about measurement, experimental tools, and channel placement. We recorded 32-channel of cap-EEG, 18-channels of ear-EEG, and 9-channels Inertial Measurement Unit (IMU) sensors. We used a wireless interface (MOVE system, Brain Product GmbH) and Ag/AgCl electrodes to acquire EEG signals from the scalp and Smarting System (mBrainTrain LLC) and cEEGrid electrodes to acquire EEG signals from ear. Three wearable IMU sensors have recorded the movement on the head, left and right ankles. The cap electrodes were placed according to the 10-20 international system at locations: Fp1, Fp2, AFz, F7, F3, Fz, F4, F8, FC5, FC1, FC2, FC6, C3, Cz, C4, CP5, CP1, CP2, CP6, P7, P3, Pz, P4, P8, PO7, PO3, POz, PO4, PO8, O1, Oz, and O2. Ear-EEG electrodes were cEEGrid, having 10 channels on left side (L1 to L10), 8 channels on right side (R1 to R8) and GND and REF in the middle of right side. The impedances were maintained below 10 $k\Omega$ for both scalp and ear-EEG. We set the sampling rate as 500 Hz for cap and ear-EEG and 128 Hz for IMU sensors. The dataset is published in 2020 International BCI Competition dataset storage (https://osf.io/pq7vb/).

All BCI experiments were developed based on the OpenBMI \cite{Lee2016OpenBMI}, BBCI \cite{krepki2007berlin} and Psychophysics toolboxes \cite{kleiner2007what}. We performed down sampling or resampling to 100 Hz for all measurement and high-pass filter using finite impulse response filter passing above 3 Hz.

\subsection{Proposed Method}
We used ensemble-based CNNs to train ERP responses described in Fig. \ref{fig2}. Ensemble networks are to combine the predictions of several base predictors (each colored block in Fig. \ref{fig2}) constructed with a given learning algorithm in order to improve generality or robustness over a single predictors. Since the ratio of target and non-target is 0.2, the non-target data was divided into four groups. Each ensemble model forwarded with one group of non-target data and target data, and then the models averaged the gradients of all four groups to update the weights of model.

The neural networks are three CNN-hidden layers and a fully-connected hidden layers. In the first layer, eight kernels, channel-wise convolution, having a size of 1 by the number of channels are used and the feature maps have a shape of time by 1. 
The feature maps are calculated by 
\begin{align}
x_k = f(\sigma_s(p))
\end{align}
where $\sigma_k$ is convolutional function, $p$ is a position of input, $f(\cdot)$ function is activation function, rectified linear unit (ReLU) is used for the activation function and is denoted by
\begin{align}
f(z) = max(0,z)
\end{align}

\begin{table*}[t]
\normalsize
\centering{
\caption{The AUC of Scalp-EEG and Ear-EEG for All Subjects}
\begin{tabular}{|c|c|c|c|c|c|c|c|c|c|c|}
\hline
\multirow{4}{*}{\begin{tabular}[c]{@{}c@{}}Scalp-\\ EEG\end{tabular}} & \textbf{Subject} & \textit{S1}  & \textit{S2}  & \textit{S3}  & \textit{S4}  & \textit{S5}  & \textit{S6}  & \textit{S7} & \textit{S8}                & \textit{S9}                \\ \cline{2-11} 
                                                                      & \textbf{AUC}     & 0.900        & 0.802        & 0.776        & 0.890        & 0.674        & 0.817        & 0.709       & 0.486                      & 0.759                      \\ \cline{2-11} 
                                                                      & \textbf{Subject} & \textit{S10} & \textit{S11} & \textit{S12} & \textit{S13} & \textit{S14} & \textit{S15} & \multicolumn{3}{c|}{\textit{Average}}                                 \\ \cline{2-11} 
                                                                      & \textbf{AUC}     & 0.597        & 0.730        & 0.716        & 0.652        & 0.618        & 0.796        & \multicolumn{3}{c|}{0.728$\pm$0.108}                     \\ \hline 
\multirow{4}{*}{\begin{tabular}[c]{@{}c@{}}Ear-\\ EEG\end{tabular}}   & \textbf{Subject} & \textit{S1}  & \textit{S2}  & \textit{S3}  & \textit{S4}  & \textit{S5}  & \textit{S6}  & \textit{S7} & \textit{S8}                & \textit{S9}                \\ \cline{2-11} 
                                                                      & \textbf{AUC}     & 0.592        & 0.669        & 0.573        & 0.663        & 0.738        & 0.574        & 0.514       & \multicolumn{1}{l|}{0.545} & \multicolumn{1}{l|}{0.731} \\ \cline{2-11} 
                                                                      & \textbf{Subject} & \textit{S10} & \textit{S11} & \textit{S12} & \textit{S13} & \textit{S14} & \textit{S15} & \multicolumn{3}{c|}{\textit{Average}}                                 \\ \cline{2-11} 
                                                                      & \textbf{AUC}     & 0.565        & 0.618        & 0.704        & 0.679        & 0.338        & 0.477        & \multicolumn{3}{c|}{0.599$\pm$0.103}                     \\ \hline 
\end{tabular}
}
\end{table*}

\begin{figure*}[ht]
    \centering
    \includegraphics{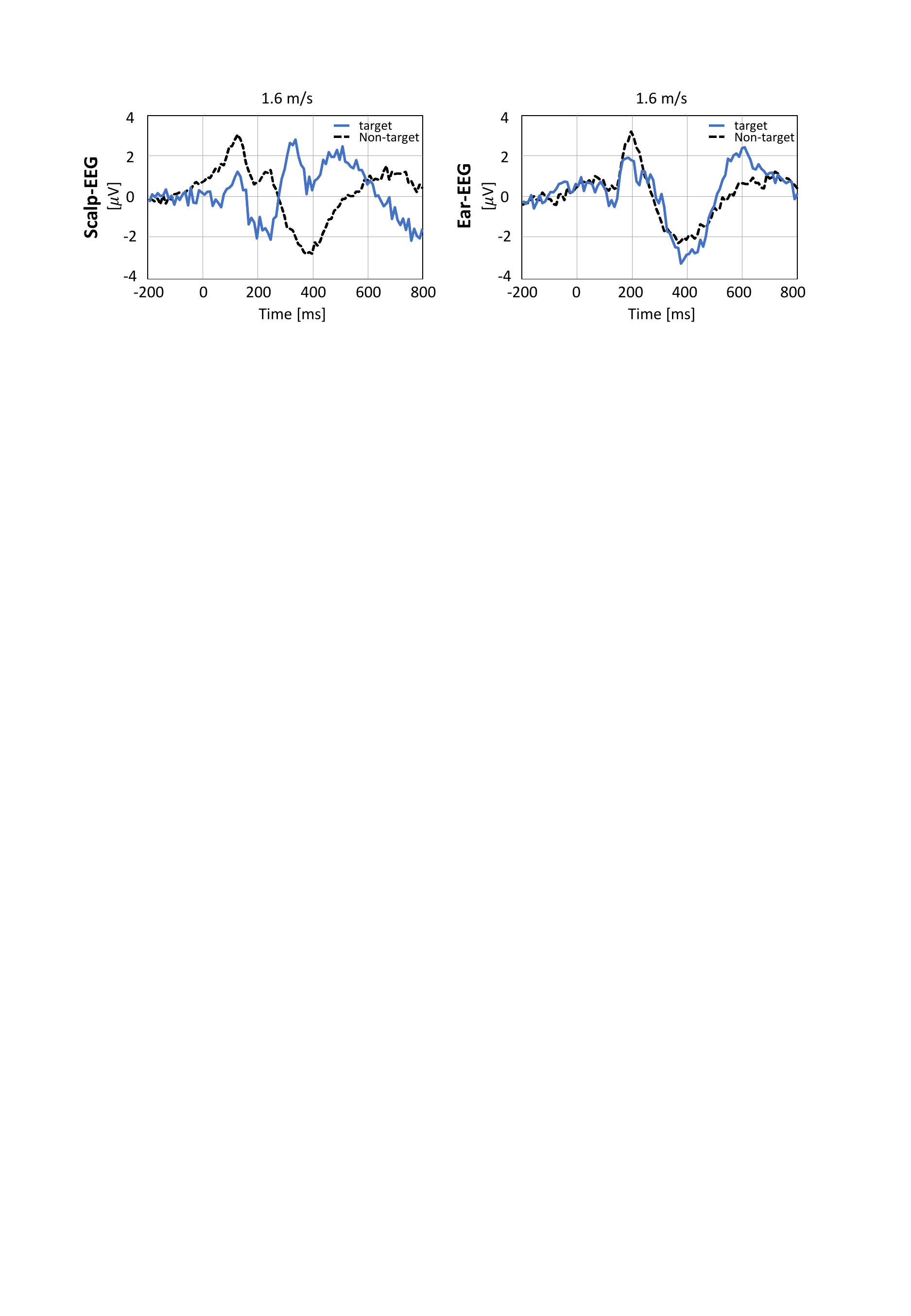}
    \caption{ERP waves of scalp- and ear-EEG at 1.6\,m/s. The plots show the grand average of the target and non-target responses indicated in blue and black line, respectively.}
    \label{fig3}
\end{figure*}

The convolutional function $\sigma_k$ can be represented by 
\begin{align}
\sigma_k (p)=b_k+\sum_{i=1}^{i=K_x} \sum_{j=1}^{j=K_y}{x^{i,j}\times w_k^{i,j}}
\end{align}
where $b_k$ is bias of kernel $k$, $K_x$ and $K_y$ is kernel size, $x$ is input matrix, and $w_k$ is weight of kernel $k$. 

Cross-entropy loss, logarithm multiplying classes, are used for the loss function of the algorithm, which is denoted by
\begin{align}
loss = -t \times log(F(s)) - (1-t) \times log(1-F(s))
\end{align}
where $t$ is a class in $[0,1]$, $s$ is the ground truth, and $F(\cdot)$ is prediction function. The learning rate was 0.01 and weights were initialized with a normal distribution. The number of epochs was 50 and batch size was 32.


\section{Results and Discussion}
For evaluating our proposed methods, we used AUC as the metric due to different number of target and non-target. The results were statistically analyzed using the statistical method of t-test. Table I shows the results of analysis for ERP from scalp and ear-EEG and indicates all subjects' performance for each method. Fig. \ref{fig3} shows the results of the ERP responses from scalp- and ear-EEG, plotting the target and non-target waves.

\subsection{Ensemble Module}
We used an ensemble module which can improve the performance when training imbalanced data. The ensemble module trained partitioned data set consisting of four non-target data set and the full target data set. The separated ensemble model forwarded the data, and then calculated the gradients of weights by averaging the gradients from all ensemble models. The total model weights were updated with all non-target and target data set at once.
The AUC of proposed method was 0.728$\pm$0.108 in average for all subject.

\subsection{ERP Responses}
Fig. \ref{fig3} shows the ERP responses of scalp and ear-EEG with the target and non-target responses. In scalp-EEG, the difference between target and non-target was huge, in particularly around at 300\,ms. Whereas the ERP responses from ear-EEG were apparent, the difference between target and non-target in ear-EEG was comparatively inferior. The ERP responses in ear-EEG showed the possibility to use practical BCI in the ambulatory environment.

\section{Conclusion}
In this study, we proposed an ensemble CNN architecture in ambulatory environment decoding visual ERP responses in scalp- and ear-EEG. As practical BCIs require a robust system in an ambulatory environment and simple hardware usable in the real-world, we show that the proposed method improved the BCI performance in the ambulatory environment.
The results for recognizing human intention in an ambulatory environment without using artifact removal methods had the reasonable performance although the data set was imbalanced. However, the performances were lower that could not fit network because of the artifacts’ variance. 
In conclusion, we showed that ensemble-based convolutional neural networks could show reasonable performance even in the ambulatory environment. However, it was difficult to increase the performance fitting from training data in standing condition to test data in walking condition due to huge artifacts features. In the future, the study removing noisy signals but remaining essential components was necessary to recognize the human intention for a different session in the ambulatory environment.

\section*{Acknowledgment}
The authors would like to thank N.-S. Kwak for his help with the design of experimental setting, M. Lee and J.-H. Jeong for their help with advising the research, and G.-H. Shin for his assistance with data collection.

\bibliographystyle{IEEEtran}
\bibliography{bibliography}

\end{document}